\documentclass[amsmath, amssymb, aps, prl, reprint, groupedaddress]{revtex4-2}

\usepackage[utf8]{inputenc} 
\usepackage[T1]{fontenc}    
\usepackage{graphicx}       
\usepackage{hyperref}       
\usepackage{booktabs}       
\usepackage{amsfonts}       
\usepackage{nicefrac}       
\usepackage{microtype}      
\usepackage{xcolor}         
\usepackage{multirow}       
\usepackage{xspace}         

\definecolor{teal}{RGB}{0, 181, 173}
\definecolor{dblue}{RGB}{13, 13, 54}
\definecolor{magenta}{RGB}{239, 71, 111}
\definecolor{gray}{RGB}{138, 139, 146}

\hypersetup{
    colorlinks=true,
    citecolor=teal,
    linkcolor=teal,
    urlcolor=teal,
}
\newcommand{\etal}{\textit{et al}.\xspace}
\newcommand{\smote}{\texttt{SMOTE}\xspace}
\newcommand{\relu}{\texttt{relu}\xspace}
\newcommand{\sigmoid}{\texttt{sigmoid}\xspace}

\begin{document}

\title{Evaluating quantum generative models via imbalanced data classification benchmarks}

\author{Graham R. Enos}
\affiliation{Rigetti Computing, 775 Heinz Ave, Berkeley, California, 94710}
\email[Corresponding author: ]{genos@rigetti.com}
\author{Matthew J. Reagor}
\affiliation{Rigetti Computing, 775 Heinz Ave, Berkeley, California, 94710}
\author{Eric Hulburd}
\affiliation{Rigetti Computing, 775 Heinz Ave, Berkeley, California, 94710}

\begin{abstract}
A limited set of tools exist for assessing whether the behavior of quantum machine learning models diverges from conventional models, outside of abstract or theoretical settings.
We present a systematic application of explainable artificial intelligence techniques to analyze synthetic data generated from a hybrid quantum-classical neural network adapted from twenty different real-world data sets, including solar flares, cardiac arrhythmia, and speech data.
Each of these data sets exhibits varying degrees of complexity and class imbalance. 
We benchmark the quantum-generated data relative to state-of-the-art methods for mitigating class imbalance for associated classification tasks.
We leverage this approach to elucidate the qualities of a problem that make it more or less likely to be amenable to a hybrid quantum-classical generative model.
\end{abstract}

\maketitle{}

\section{Introduction}
Quantum machine learning has attracted significant attention in recent years owing to its potential for realizing an exponentially larger vector space for processing~\cite{Schuld_2014, Biamonte2017}. Generative modeling is a particularly attractive task for current quantum hardware platforms, where quantum bits (qubits) can be used as an exponentially large, trainable latent space and sampled as stochastic inputs to traditional CPU/GPU frameworks as a hybrid model.
Understanding these new techniques from first-principles theory remains a daunting task~\cite{Schuld2022}, leading to heuristic-based evaluation for possible quantum advantage, as developed by Hibat-Allah \etal \cite{Hibatallah2023}.
A key challenge has been the introduction of methods amenable to a broad collection of real-world data that shed light on how quantum-generated data influence the hybrid model results.

Here, we demonstrate the use of a hybrid quantum-classical generative modeling technique known as a Quantum Circuit Associative Adversarial Network (QC-ANN), first introduced by Rudolph \etal \cite{Zapata2022}, to create synthetic data for training a classification model.
We compare the performance of this approach to purely classical methods over a diverse collection of imbalanced data sets, employing explainable artificial intelligence tools to determine what features of a data set make it more or less likely to be amenable to the hybrid quantum-classical approach versus purely classical techniques.
We observe that the relative performance of the QC-ANN model is best with very few original training samples in the minority case, for example, finding statistically significant improved classification outcomes for a financial use-case of bankruptcy prediction with a 30:1 class imbalance.

\begin{figure*}[htbp]
    \centering
    \includegraphics[width=\textwidth]{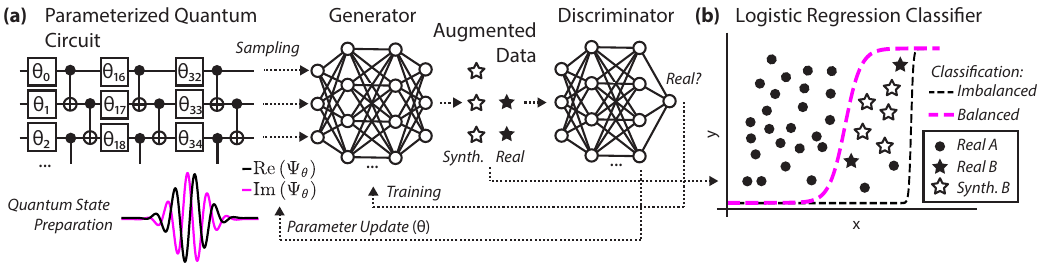}
    \caption{Quantum generative model and benchmarks. \textbf{(a)} An architecture for Quantum Circuit Adversarial Neural Network (QC-AAN) which uses a parameterized quantum circuit (left) as a high-dimensional, trainable noise source for a Generative Adversarial Network (GAN) (right). \textbf{(b)} A collection of twenty imbalanced classification tasks is used to interpret the output of the QC-AAN.}
    \label{fig:architecture}
\end{figure*}

Methods to ameliorate imbalance and successfully train classification models are important research areas on their own.
With a large enough imbalance, a model will often ``memorize'' the over-represented class and not make predictions for the underrepresented class(es).
Indeed, classification problems with imbalanced data regularly arise in real-world scenarios such as fraud detection, medical imaging, computer and network security, and environmental applications~\cite{Johnson2019}.
The techniques developed here may be directly adapted to these settings.

\section{Related Work}

When quantum machine learning was a more nascent research area, experiments focused on smaller data sets, such as the ``bars-and-stripes'' \cite{Zhu2019} dataset.
As it further develops, quantum machine learning experiments are shifting their focus to larger, more real-world data sets, such as handwritten digits \cite{Zapata2022}, correlated currency pairs \cite{Coyle2021}, medical imagery \cite{Landman2022}, and synthetic weather radar \cite{Enos2021}.

Recently, Hibat-Allah \etal \cite{Hibatallah2023} built upon metrics for generalization introduced by Gili \etal \cite{Gili2022evaluating} to propose a quantitative framework for quantum advantage for generative modeling.
The focus on generalization is well-motivated in this study because of the mathematical properties of the specific, artificial distributions the generative models are tasked with representing.
While this approach leads to some interesting observations---such as that the hybrid model performs well in the low data regime, in practice, benchmarking generalization is notoriously difficult in real-world data.
Indeed, for traditional machine learning models, there are other metrics such as the Inception Score \cite{Salimans2016}, Frech\'et Inception Distance \cite{Heusel2018}, and Precision and Recall \cite{Simon2019} that are used to measure a generative model's ability to create high-quality synthetic data, but they all fall short in examining how well a model generalizes.

Focusing on downstream classification outcomes provides generic, practical evaluation criteria.
In that direction, within a clinical data context, Kazdaghli \etal \cite{Kazdaghli2023} have recently evaluated classification outcomes using quantum-generated data to impute missing features.
Their approach leverages sampling methods that may have an asymptotic advantage over classical algorithms on future error-corrected quantum hardware.
Unfortunately, these methods are not well-suited for benchmarking existing systems with limited capacity for entanglement.

\section{Methods}
\subsection{Hybrid quantum-classical generative model}

First introduced in 2014 \cite{Goodfellow2014}, a \textit{generative adversarial network} (GAN) consists of two neural networks: a \textit{generator} which takes in (typically Gaussian) random noise and creates synthetic data that resembles some training data, and a \textit{discriminator} that discerns true training data from synthetic data created by the generator.
These two networks are trained in tandem until the generator can create synthetic data of sufficient quality.
Arici and Celikyilmaz\cite{Arici2016} introduced an extension called an \textit{associative adversarial network} (AAN).
Instead of the generator taking in pseudorandom noise, the generator and discriminator share a smaller generative model that adapts to the state of the overall network.
During training, the inner model is tuned to capture the latent space of the discriminator; this means that the generator receives input that more closely resembles the underlying distribution that the discriminator is looking for, speeding up the training of the overall network.

While the original AAN setup used a classical Restricted Boltzmann machine (RBM) for the 
 model, Rudolph \etal \cite{Zapata2022} introduced a Quantum Circuit Associative Adversarial Network (QC-AAN) which replaces that inner tunable noise generator with a Quantum Circuit Born Machine (QCBM) \cite{Benedetti2019}.
During training, samples from a quantum processing unit (QPU) running the QCBM are used as input to the generator.
Periodically, the generator-discriminator training loop is paused, and the QCBM is trained for several iterations (using Sinkhorn divergence as in Coyle \etal \cite{Coyle2021}) to output bitstrings that more closely resemble a binomial distribution parameterized by the discriminator's penultimate layer.
We used \texttt{TensorFlow} \cite{tensorflow2015} for defining, training, and evaluating the neural networks.
For the two experiments described in this note, we implemented QC-AANs using \texttt{pyQuil} \cite{Smith2016} for defining and training the QCBM, connecting to Rigetti Computing's quantum virtual machines (QVMs) and QPUs \cite{Karalekas2020}.

\subsection{First experiment: a comparison across data sets}

Keeping with the metaphor of Hibat-Allah \cite{Hibatallah2023}, for our first experiment we ran multiple races between purely classical and the QC-AAN approach across several data sets, to better understand what characteristics of an imbalanced data set may make it a good candidate---or not---for this hybrid quantum-classical approach.
We turned to the catalog of data sets from Zejin \cite{Zejin2011} as made available by the \texttt{imbalanced-learn} software package \cite{imblearn2017}, filtering to the 19 data sets with 16 or more features to make it worthwhile to model them on a QPU.
In addition to these (largely bioinformatics-related) sets, we incorporated the company bankruptcy data set first examined by Liang \etal \cite{Liang2016}, available in the UCI Machine Learning Repository \cite{UCI2020}.
This data set contains 94 financially relevant features, such as total liability/equity ratio or effective tax rate, and a label indicating whether each company went bankrupt or not for 6,819 companies---of which only 220 are in the positive class.
We put all data sets through minimal preprocessing: scaling all features to the $[0, 1]$ range and breaking them into a 75\%$\left.\middle/\right.$25\% train/test split.

We next set up a QC-AAN implementation tailored to the characteristics of each data set to adequately transform and reshape the data to appropriate dimensions and distributions.
In particular, let $f$ be the number of features of a particular data set $d$, and let $q$ be the number of qubits in our QCBM; by design, $f > q$.
We construct intermediate fully-connected layers by repeatedly doubling $q$ until it is no longer less than $f$, taking \relu activation; for notation's sake, suppose there are $n$ such layers.
Then the architecture of the neural networks was as follows:

\begin{itemize}
    \item generator:
    \begin{enumerate}
        \item[0.] input layer with dimension equal to sampler's latent dimension $q$, the number of qubits;
        \item[$k = 1\dots n$.] fully-connected layer of dimension $2kq$ with \relu activation;
        \item[$n + 1$.] fully-connected output layer with dimension $f$ and \relu activation.
    \end{enumerate}
    \item discriminator:
    \begin{enumerate}
        \item[0.] input layer with dimension equal to $f$, the number of features;
        \item[$k = 1\dots n$.] fully-connected layer of dimension $2(n - k + 1)q$ with \relu activation;
        \item[$n + 1$.] fully-connected layer with latent dimension $q$; and
        \item[$n + 2$.] fully-connected output layer with dimension 1 and \sigmoid activation.
    \end{enumerate}
\end{itemize}

We trained the neural networks with the \textit{Adam} \cite{Kingma2014} optimizer, periodically stopping to train the QCBM as previously mentioned.
We opted for a simple final classifier---logistic regression from \texttt{scikit-learn} \cite{scikit-learn2011}---deliberately chosen to highlight the different effects of the different approaches.
Each data set $d$ had an underrepresented positive class, each with varying ratios of negative to positive class, number of features, and other characteristics summarized in Table~\ref{tab:datasets} in the appendix.

\subsection{Second experiment: focusing on company bankruptcy}

For our second experiment, we explored the company bankruptcy data in more detail.
To examine how the models scaled with feature space dimension, we performed PCA on the company bankruptcy data to get $f = q$.
We deliberately opted for relatively simple neural network architectures for the generator and discriminator for this experiment to place more emphasis on the quantum portion of the hybrid algorithm, which was a 16-qubit QCBM from a 40-qubit processor.
The AAN architecture was as follows:

\begin{itemize}
    \item generator:
    \begin{enumerate}
        \item[0.] input layer with same dimension $q$ as noise source; and
        \item[1.] fully connected output layer with the same dimension $f$ as the data set and \relu activation.
    \end{enumerate}
    \item discriminator:
        \begin{enumerate}
            \item[0.] input layer with same dimension $f$ as data set;
            \item[1.] fully-connected layer with same dimension $q$ as noise source with \relu activation; and
            \item[2.] fully connected output layer with a single neuron and \sigmoid activation for prediction.
        \end{enumerate}
\end{itemize}

We trained the neural networks with the \textit{Adam} \cite{Kingma2014} optimizer, periodically stopping to train the QCBM as previously mentioned.
Both \cite{Zapata2022} and, e.g., Coyle \etal \cite{Coyle2021} have demonstrated the ability of QCBMs to generate features that induce larger discriminator error than RBMs in contexts of relatively high feature precision and trained circuit entanglement; as such, instead of comparing a QC-AAN to an AAN with an RBM as the internal model, we trained a classical GAN counterpart to this QC-AAN for comparison.
It had the same generator and discriminator setup, differing only in the ``noise source,'' which was standard pseudorandom Gaussian noise.

\section{Experimental Results}

For both experiments, we collected generated data from the trained QC-AANs and their classical counterparts to examine via training follow-on classifiers and using explainable AI tools.

\subsection{First experiment}

For each data set $d$, we trained a logistic regression classifier on $d$ with

\begin{enumerate}
    \item no additional data;
    \item the positive class upsampled at random;
    \item the positive class inflated via the \smote algorithm \cite{Chawla2002} as implemented in \texttt{imblearn} \cite{imblearn2017}; and
    \item the positive class augmented with QC-AAN-generated data using a quantum circuit with 16 qubits.
\end{enumerate}

\begin{figure*}[htbp]
    \centering
    \includegraphics[width=\textwidth]{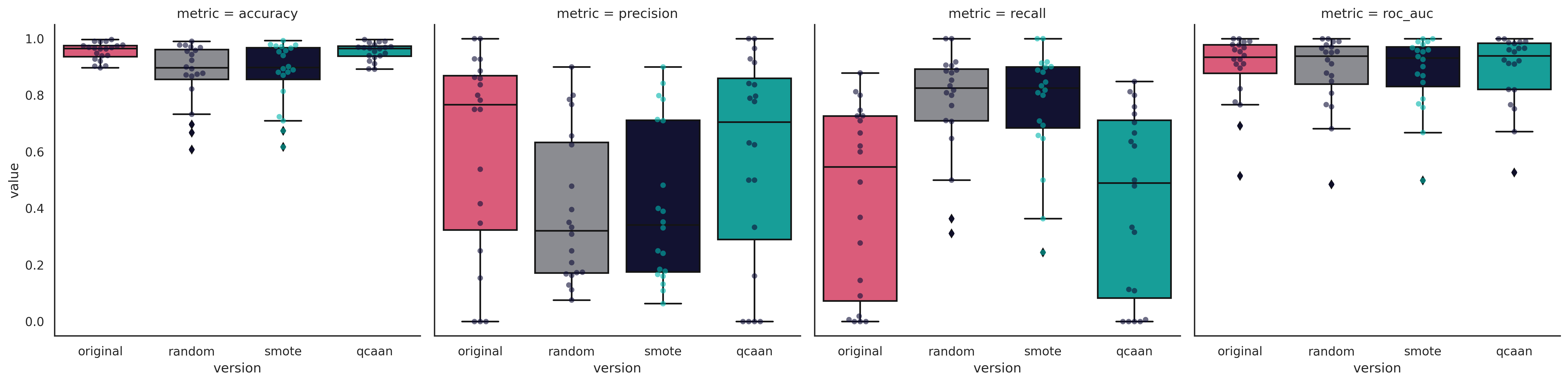}
    \caption{Classification metrics across data sets, comparing \textcolor{magenta}{original data}, \textcolor{gray}{randomly} resampled, resampled with \textcolor{dblue}{\smote}, and synthesized via a 16 qubit \textcolor{teal}{QC-AAN}.}
    \label{fig:catalog/boxplot}
\end{figure*}

For each of these approaches, we trained our final logistic regression classifier and calculated four metrics of classification quality on a holdout test set, viz. accuracy, precision, recall, and area under the ROC curve.
During development, we also ran an 8-qubit QC-AAN simulated on a noiseless quantum circuit simulator. The aggregation of all these metrics across data sets is summarized in Figure~\ref{fig:catalog/boxplot}.

\begin{figure*}[htbp]
    \centering
    \includegraphics[width=0.75\textwidth]{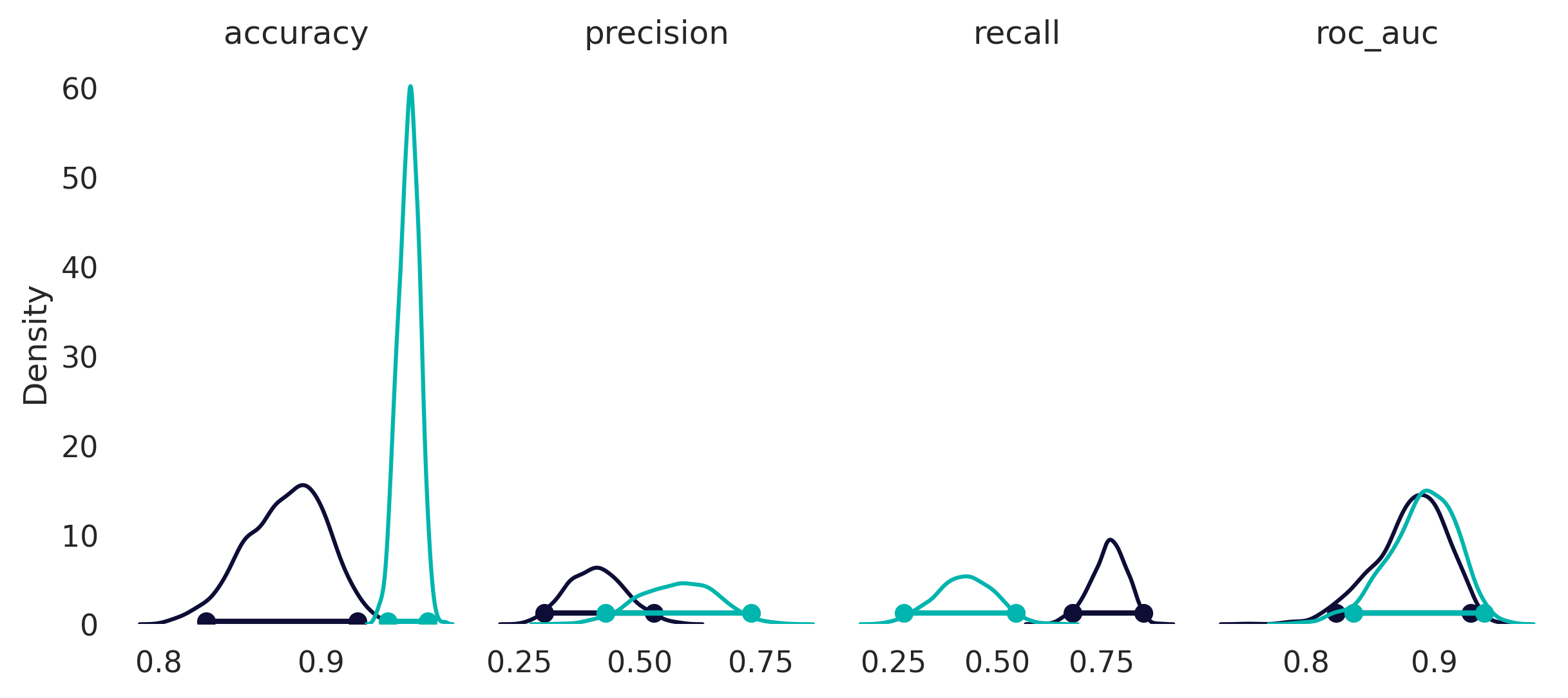}
    \caption{Bayesian bootstrapped metrics with 95\% HDIs, \textcolor{dblue}{\smote} vs. 16 qubit \textcolor{teal}{QC-AAN}.}
    \label{fig:catalog/bayesboot}
\end{figure*}

To better understand the difference in classification metrics, we first turned to post hoc pairwise testing.
Dunn's test \cite{Dunn1964} as implemented in \texttt{scikit-posthocs} \cite{scikit-posthocs2019} highlighted only accuracy and recall as having statistically significant ($\alpha = 0.05$) differences between classical and quantum approaches.
However, a visual examination of Figure~\ref{fig:catalog/boxplot} suggested there may be a difference in precision as well.
Turning to a different statistical approach, we reach for the Bayesian bootstrap \cite{Rubin1981}; Bayesian bootstrapped means with 95\% highest density intervals of four classification metrics (1,000 replications) suggested that precision is also worth examining in addition to accuracy and recall, as shown in Figure~\ref{fig:catalog/bayesboot}.

\begin{figure*}
    \centering
    \includegraphics[width=0.5\textwidth]{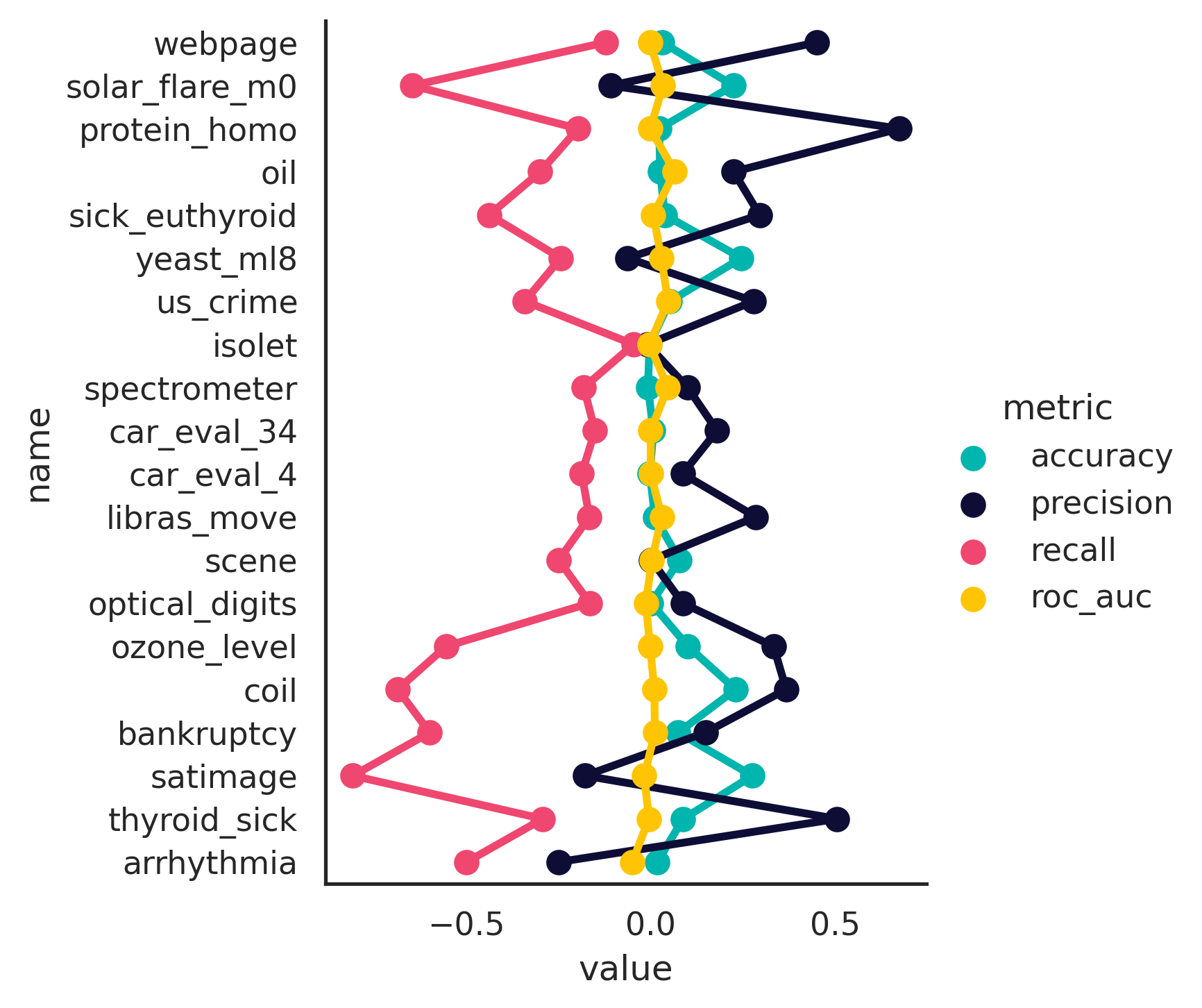}
    \caption{Differences of classification metrics, \smote values subtracted from QC-AAN values.}
    \label{fig:catalog/diffs}
\end{figure*}

To understand the features of a data set that make it more or less amenable to this hybrid quantum-classical approach, we reached for some explainable AI tools.
For each of the three classification metrics highlighted as worth exploring by our Bayesian bootstrap methodology, we calculated the difference between the \smote value and the one from the QC-AAN on the quantum processor; the values of these diffs of metrics can be viewed in Figure~\ref{fig:catalog/diffs}.
We next treat that value across the 19 data sets as a target variable we wished to predict from the characteristics of the data sets in question.
In addition to the features available directly from \cite{imblearn2017} in Table~\ref{tab:datasets} in the appendix, we also computed some additional distance-based ones.
For both the negative class $n$ and positive class $p$, we computed the minimum, median, and maximum pairwise distance between points of the same class.
In addition, we computed the minimum, median, and maximum of the pairwise distances between points in $n$ and points in $p$.
While these additional features are correlated (a larger minimum implies a larger median, etc.), we wanted to examine as much metadata about the data sets in question as we could, letting a more sophisticated model decide what was relevant.
See Table~\ref{tab:meta} in the appendix for the metadata values.

Then for each metric, we trained an \texttt{XGBoost} \cite{Chen2016} regressor on the data set features and metadata, trying to predict the value of that metric.
Once completed, we calculated model aspect importance and ceteris paribus profiles via \texttt{dalex} \cite{dalex2021} to determine the most relevant features and better understand their influence on the particular metric.
We summarize the findings as follows:

\begin{itemize}
    \item[\textbf{Accuracy}] The QC-AAN looks uniformly the same or better than that of \smote, as one can see from Figure~\ref{fig:catalog/diffs}.
    However, as the distance between points in the same group ($\mathrm{d}_n$ and $\mathrm{d}_p$ respectively) increases, the performance from the \smote-generated data gains on the quantum-generated data.
    \item[\textbf{Precision}] The performance influenced by the QC-AAN as implemented on quantum hardware often meets or exceeds that from the \smote implementation; it especially outperforms its classical contender as the ratio of negative to positive increases, i.e. as the problem becomes more imbalanced.
    \item[\textbf{Recall}] This is the one metric where the classical \smote approach almost uniformly outperformed the hybrid quantum-classical implementation.
    However, the quantum approach slightly gains on the classical \smote setup as the distance between positive samples ($\mathrm{d}_p$) increases.
\end{itemize}

\subsection{Second experiment}

Once the classical GAN and hybrid QC-AAN were trained, we first generated several samples from both to compare and contrast them.
A simple logistic regression (implemented via \texttt{scikit-learn} \cite{scikit-learn2011}) could easily tell the difference between the outputs of the two different generative models, as we can see from Table~\ref{tab:distinguish}.
This suggests that the hybrid quantum-classical approach was doing something unique---generating linearly distinguishable features that are not as easily attainable through classical generative means---rather than just generating the same data we could have attained classically.

\begin{table*}[htbp]
    \centering
    \caption{Confusion matrix for logistic regression distinguishing classical from quantum generated data (16 qubits on quantum hardware).}
    \label{tab:distinguish}
    \begin{tabular}{@{}cc cc@{}}
        \multicolumn{1}{c}{} &\multicolumn{1}{c}{} &\multicolumn{2}{c}{Predicted} \\ 
        \cmidrule{3-4}
        \multicolumn{1}{c}{} &
            \multicolumn{1}{c}{} & 
            \multicolumn{1}{c}{Classical} & 
            \multicolumn{1}{c}{Quantum} \\[1.125ex] 
        \cline{2-4}\\[0.1ex]
        \multirow[c]{2}{*}{\rotatebox[origin=tr]{90}{Actual}}
            & Classical  & 1300 & 0   \\[1.5ex]
        & Quantum  & 1   & 1251 \\ 
        \cline{2-4}
    \end{tabular}
\end{table*}

\begin{figure*}
    \centering
    \includegraphics[width=0.7\textwidth]{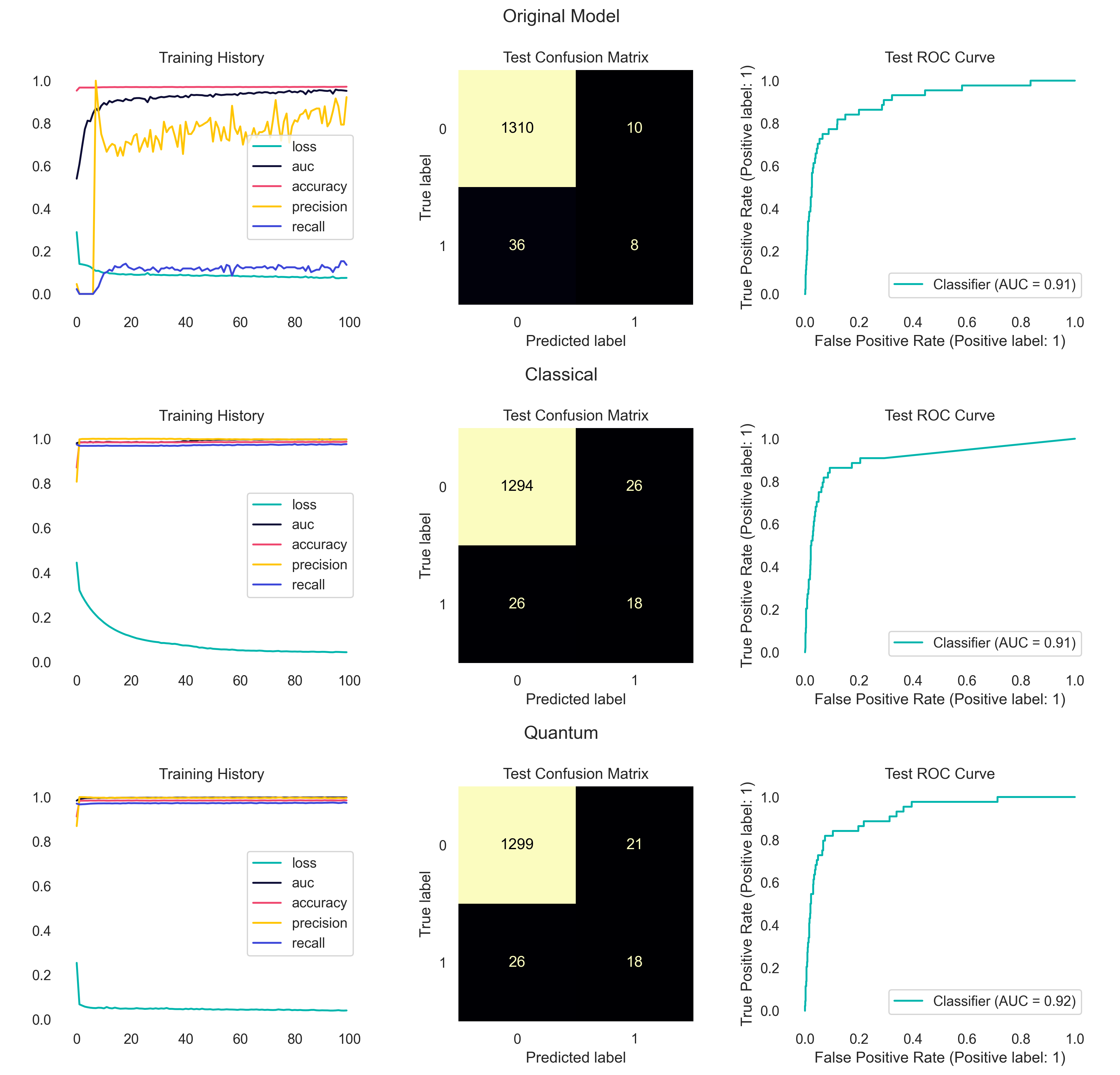}
    \caption{Training and prediction outcomes with the original data set only, original plus classical GAN-generated data, and original plus QC-AAN-generated data (quantum hardware, 16 qubits).}
    \label{fig:bankruptcy/approaches}
\end{figure*}

For all three approaches---using original data only, using original data plus GAN-generated synthetic data, and using original data plus QC-AAN-generated synthetic data---we trained a feed-forward neural network consisting of fully-connected layers of dimension 64, 32, 16, and 4, each with \relu activation, and a final output layer of a single node with \sigmoid; this was an elaboration of the original multilayer perceptron used in \cite{Liang2016}.
As evidenced by Figure~\ref{fig:bankruptcy/approaches}, using synthetic data from classical GAN improves training and prediction compared to using only the original data; QC-AAN (16 qubits) improves even more.
The loss curve improves, dropping to zero sooner with the QC-AAN-generated data, and the other training metrics also improve, exhibiting even smaller variance with the QC-AAN-generated data than the GAN-generated.
Moreover, final classification performance on the hold-out test set improved; final classification precision was greater with the QC-AAN-generated data than the original or GAN-generated data, as is further illustrated by the confusion matrices and ROC curves in Figure~\ref{fig:bankruptcy/approaches}.

\section{Discussion}

We have introduced a framework for assessing quantum-generated data using classification outcomes on imbalanced data sets as a measure of synthetic data quality.
We have applied this methodology to benchmark the performance of a hybrid quantum-classical generative model known as a QC-AAN as the source of synthetic data.
In the first experiment, we shed light on some characteristics of imbalanced data sets that might make them more or less amenable to this hybrid quantum-classical approach compared to modern classical methods.
In our second, we demonstrated improved classification performance over training with the original data and training with additional data generated by an analogous purely classical GAN in a financial application predicting potential business bankruptcies with very few training samples in the positive class compared to the overrepresented negative one.
Taken together, these make a strong case for quantum generative modeling as an approach to ameliorating class imbalance in classification problems, especially in applications that prioritize high precision over other binary classification metrics.

In future work, we plan to evaluate how these findings scale as the size of the quantum computational resource grows, including the number of qubits and degree of entanglement.
It will also be important to further understand the interplay between the quantum ``noise generator'' and the classical AAN around it.
For instance, with more qubits, it is possible to generate data from larger-dimensional latent distributions, potentially cutting back on the need for complicated classical network layers. On the other hand, those classical networks may need to grow in complexity as well to adequately handle, shape, and transform the more complicated generated data coming from the QPU.
In these efforts, it will be necessary to contrast more sophisticated models with ever increasingly advanced classical counterparts like \smote, GANs, and AANs with the original RBM setup, at varying levels of resampling to examine how synthetic data quality fares as a function of the number of generated samples.

\begin{acknowledgments}
All quantum computations were run via Rigetti Computing's \texttt{pyQuil} \cite{Smith2016} package, connecting to Rigetti Computing's QVMs and QPUs \cite{Karalekas2020}.
As such, the authors would like to thank the entire Rigetti team for helpful discussions, suggestions, and assistance; in particular, we thank Mark Skilbeck and Tom Elliff-O'Shea for enabling this work with extensive systems engineering support, and Karina Castano Munoz and Eric Ostby for project guidance.
Classical computations---including training and testing the neural networks and plotting final results---utilized an Ampere\textsuperscript{\textregistered} Altra\textsuperscript{\textregistered} Max\footnote{\url{https://amperecomputing.com/processors/ampere-altra}} multi-core processor.
We  thank Ampere for the use of this machine.
\end{acknowledgments}

\bibliographystyle{unsrt}
\bibliography{main}

\begin{thebibliography}{10}

\bibitem{Schuld_2014}
Maria Schuld, Ilya Sinayskiy, and Francesco Petruccione.
\newblock An introduction to quantum machine learning.
\newblock {\em Contemporary Physics}, 56(2):172--185, oct 2014.

\bibitem{Biamonte2017}
Jacob Biamonte, Peter Wittek, Nicola Pancotti, Patrick Rebentrost, Nathan
  Wiebe, and Seth Lloyd.
\newblock Quantum machine learning.
\newblock {\em Nature}, 549(7671):195--202, 2017.

\bibitem{Schuld2022}
Maria Schuld and Nathan Killoran.
\newblock Is quantum advantage the right goal for quantum machine learning?
\newblock {\em PRX Quantum}, 3:030101, Jul 2022.

\bibitem{Hibatallah2023}
Mohamed Hibat-Allah, Marta Mauri, Juan Carrasquilla, and Alejandro
  Perdomo-Ortiz.
\newblock A framework for demonstrating practical quantum advantage: Racing
  quantum against classical generative models, 2023.

\bibitem{Zapata2022}
Manuel~S. Rudolph, Ntwali~Bashige Toussaint, Amara Katabarwa, Sonika Johri,
  Borja Peropadre, and Alejandro Perdomo-Ortiz.
\newblock Generation of high-resolution handwritten digits with an ion-trap
  quantum computer.
\newblock {\em Phys. Rev. X}, 12:031010, Jul 2022.

\bibitem{Johnson2019}
Justin~M. Johnson and Taghi~M. Khoshgoftaar.
\newblock Survey on deep learning with class imbalance.
\newblock {\em Journal of Big Data}, 6(1):27, 2019.

\bibitem{Zhu2019}
D.~Zhu, N.~M. Linke, M.~Benedetti, K.~A. Landsman, N.~H. Nguyen, C.~H.
  Alderete, A.~Perdomo-Ortiz, N.~Korda, A.~Garfoot, C.~Brecque, L.~Egan,
  O.~Perdomo, and C.~Monroe.
\newblock Training of quantum circuits on a hybrid quantum computer.
\newblock {\em Science Advances}, 5(10):eaaw9918, 2019.

\bibitem{Coyle2021}
Brian Coyle, Maxwell Henderson, Justin Chan~Jin Le, Niraj Kumar, Marco Paini,
  and Elham Kashefi.
\newblock Quantum versus classical generative modelling in finance.
\newblock {\em Quantum Science and Technology}, 6(2):024013, April 2021.

\bibitem{Landman2022}
Jonas Landman, Natansh Mathur, Yun~Yvonna Li, Martin Strahm, Skander Kazdaghli,
  Anupam Prakash, and Iordanis Kerenidis.
\newblock Quantum {M}ethods for {N}eural {N}etworks and {A}pplication to
  {M}edical {I}mage {C}lassification.
\newblock {\em {Quantum}}, 6:881, December 2022.

\bibitem{Enos2021}
Graham~R. Enos, Matthew~J. Reagor, Maxwell~P. Henderson, Christina Young, Kyle
  Horton, Mandy Birch, and Chad Rigetti.
\newblock Synthetic weather radar using hybrid quantum-classical machine
  learning, 2021.

\bibitem{Gili2022evaluating}
Kaitlin Gili, Marta Mauri, and Alejandro Perdomo-Ortiz.
\newblock Evaluating generalization in classical and quantum generative models,
  2022.

\bibitem{Salimans2016}
Tim Salimans, Ian Goodfellow, Wojciech Zaremba, Vicki Cheung, Alec Radford,
  Xi~Chen, and Xi~Chen.
\newblock Improved techniques for training gans.
\newblock In D.~Lee, M.~Sugiyama, U.~Luxburg, I.~Guyon, and R.~Garnett,
  editors, {\em Advances in Neural Information Processing Systems}, volume~29.
  Curran Associates, Inc., 2016.

\bibitem{Heusel2018}
Martin Heusel, Hubert Ramsauer, Thomas Unterthiner, Bernhard Nessler, and Sepp
  Hochreiter.
\newblock Gans trained by a two time-scale update rule converge to a local nash
  equilibrium, 2018.

\bibitem{Simon2019}
Loïc Simon, Ryan Webster, and Julien Rabin.
\newblock Revisiting precision and recall definition for generative model
  evaluation, 2019.

\bibitem{Kazdaghli2023}
Skander Kazdaghli, Iordanis Kerenidis, Jens Kieckbusch, and Philip Teare.
\newblock Improved clinical data imputation via classical and quantum
  determinantal point processes, 2023.

\bibitem{Goodfellow2014}
Ian Goodfellow, Jean Pouget-Abadie, Mehdi Mirza, Bing Xu, David Warde-Farley,
  Sherjil Ozair, Aaron Courville, and Yoshua Bengio.
\newblock Generative adversarial nets.
\newblock In {\em Advances in neural information processing systems}, pages
  2672--2680, 2014.

\bibitem{Arici2016}
Tarik Arici and Asli Celikyilmaz.
\newblock Associative adversarial networks.
\newblock {\em CoRR}, abs/1611.06953, 2016.

\bibitem{Benedetti2019}
Marcello Benedetti, Delfina Garcia-Pintos, Oscar Perdomo, Vicente
  Leyton-Ortega, Yunseong Nam, and Alejandro Perdomo-Ortiz.
\newblock A generative modeling approach for benchmarking and training shallow
  quantum circuits.
\newblock {\em npj Quantum Information}, 5(1):45, 2019.

\bibitem{tensorflow2015}
Mart\'{i}n Abadi, Ashish Agarwal, Paul Barham, Eugene Brevdo, Zhifeng Chen,
  Craig Citro, Greg~S. Corrado, Andy Davis, Jeffrey Dean, Matthieu Devin,
  Sanjay Ghemawat, Ian Goodfellow, Andrew Harp, Geoffrey Irving, Michael Isard,
  Yangqing Jia, Rafal Jozefowicz, Lukasz Kaiser, Manjunath Kudlur, Josh
  Levenberg, Dandelion Man\'{e}, Rajat Monga, Sherry Moore, Derek Murray, Chris
  Olah, Mike Schuster, Jonathon Shlens, Benoit Steiner, Ilya Sutskever, Kunal
  Talwar, Paul Tucker, Vincent Vanhoucke, Vijay Vasudevan, Fernanda Vi\'{e}gas,
  Oriol Vinyals, Pete Warden, Martin Wattenberg, Martin Wicke, Yuan Yu, and
  Xiaoqiang Zheng.
\newblock {TensorFlow}: Large-scale machine learning on heterogeneous systems,
  2015.
\newblock Software available from tensorflow.org.

\bibitem{Smith2016}
Robert~S. Smith, Michael~J. Curtis, and William~J. Zeng.
\newblock A practical quantum instruction set architecture, 2016.

\bibitem{Karalekas2020}
Peter~J Karalekas, Nikolas~A Tezak, Eric~C Peterson, Colm~A Ryan, Marcus~P
  da~Silva, and Robert~S Smith.
\newblock A quantum-classical cloud platform optimized for variational hybrid
  algorithms.
\newblock {\em Quantum Science and Technology}, 5(2):024003, apr 2020.

\bibitem{Zejin2011}
Zejin Ding.
\newblock {\em Diversified Ensemble Classifiers for Highly Imbalanced Data
  Learning and Its Application in Bioinformatics}.
\newblock PhD thesis, Georgia State University, 2011.

\bibitem{imblearn2017}
Guillaume Lema{{\^i}}tre, Fernando Nogueira, and Christos~K. Aridas.
\newblock Imbalanced-learn: A python toolbox to tackle the curse of imbalanced
  datasets in machine learning.
\newblock {\em Journal of Machine Learning Research}, 18(17):1--5, 2017.

\bibitem{Liang2016}
Deron Liang, Chia-Chi Lu, Chih-Fong Tsai, and Guan-An Shih.
\newblock Financial ratios and corporate governance indicators in bankruptcy
  prediction: A comprehensive study.
\newblock {\em European Journal of Operational Research}, 252(2):561--572,
  2016.

\bibitem{UCI2020}
{Taiwanese Bankruptcy Prediction}.
\newblock UCI Machine Learning Repository, 2020.

\bibitem{Kingma2014}
Diederik~P. Kingma and Jimmy Ba.
\newblock Adam: {A} method for stochastic optimization.
\newblock In Yoshua Bengio and Yann LeCun, editors, {\em 3rd International
  Conference on Learning Representations, {ICLR} 2015, San Diego, CA, USA, May
  7-9, 2015, Conference Track Proceedings}, 2015.

\bibitem{scikit-learn2011}
F.~Pedregosa, G.~Varoquaux, A.~Gramfort, V.~Michel, B.~Thirion, O.~Grisel,
  M.~Blondel, P.~Prettenhofer, R.~Weiss, V.~Dubourg, J.~Vanderplas, A.~Passos,
  D.~Cournapeau, M.~Brucher, M.~Perrot, and E.~Duchesnay.
\newblock Scikit-learn: Machine learning in {P}ython.
\newblock {\em Journal of Machine Learning Research}, 12:2825--2830, 2011.

\bibitem{Chawla2002}
N.~V. Chawla, K.~W. Bowyer, L.~O. Hall, and W.~P. Kegelmeyer.
\newblock {SMOTE}: Synthetic minority over-sampling technique.
\newblock {\em Journal of Artificial Intelligence Research}, 16:321--357, jun
  2002.

\bibitem{Dunn1964}
Olive~Jean Dunn.
\newblock Multiple comparisons using rank sums.
\newblock {\em Technometrics}, 6(3):241--252, 1964.

\bibitem{scikit-posthocs2019}
Maksim Terpilowski.
\newblock scikit-posthocs: Pairwise multiple comparison tests in python.
\newblock {\em The Journal of Open Source Software}, 4(36):1169, 2019.

\bibitem{Rubin1981}
Donald~B. Rubin.
\newblock {The Bayesian Bootstrap}.
\newblock {\em The Annals of Statistics}, 9(1):130 -- 134, 1981.

\bibitem{Chen2016}
Tianqi Chen and Carlos Guestrin.
\newblock {XGBoost}: A scalable tree boosting system.
\newblock In {\em Proceedings of the 22nd ACM SIGKDD International Conference
  on Knowledge Discovery and Data Mining}, KDD '16, pages 785--794, New York,
  NY, USA, 2016. ACM.

\bibitem{dalex2021}
Hubert Baniecki, Wojciech Kretowicz, Piotr Piatyszek, Jakub Wisniewski, and
  Przemyslaw Biecek.
\newblock dalex: Responsible machine learning with interactive explainability
  and fairness in python.
\newblock {\em Journal of Machine Learning Research}, 22(214):1--7, 2021.

\bibitem{Note1}
\protect \url {https://amperecomputing.com/processors/ampere-altra}.

\end{thebibliography}

\newpage
\appendix

\begin{table*}[htbp]
    \caption{Imbalanced data sets from \cite{Zejin2011} \& \cite{imblearn2017}.}
    \label{tab:datasets}
    \begin{tabular}{lrrrrr}
        \toprule
Name & \# Features & \# Samples & \# Negative & \# Positive & Ratio \\
        \midrule
\texttt{arrhythmia}       & 278 & 452    & 427    & 25   & 17.08  \\
\texttt{car\_eval\_34}    & 21  & 1728   & 1594   & 134  & 11.90  \\
\texttt{car\_eval\_4}     & 21  & 1728   & 1663   & 65   & 25.58  \\
\texttt{coil\_2000}       & 85  & 9822   & 9236   & 586  & 15.76  \\
\texttt{isolet}           & 617 & 7797   & 7197   & 600  & 12.00  \\
\texttt{libras\_move}     & 90  & 360    & 336    & 24   & 14.00  \\
\texttt{oil}              & 49  & 937    & 896    & 41   & 21.85  \\
\texttt{optical\_digits}  & 64  & 5620   & 5066   & 554  & 9.14   \\
\texttt{ozone\_level}     & 72  & 2536   & 2463   & 73   & 33.74  \\
\texttt{protein\_homo}    & 74  & 145751 & 144455 & 1296 & 111.46 \\
\texttt{satimage}         & 36  & 6435   & 5809   & 626  & 9.28   \\
\texttt{scene}            & 294 & 2407   & 2230   & 177  & 12.60  \\
\texttt{sick\_euthyroid}  & 42  & 3163   & 2870   & 293  & 9.80   \\
\texttt{solar\_flare\_m0} & 32  & 1389   & 1321   & 68   & 19.43  \\
\texttt{spectrometer}     & 93  & 531    & 486    & 45   & 10.80  \\
\texttt{thyroid\_sick}    & 52  & 3772   & 3541   & 231  & 15.33  \\
\texttt{us\_crime}        & 100 & 1994   & 1844   & 150  & 12.29  \\
\texttt{webpage}          & 300 & 34780  & 33799  & 981  & 34.45  \\
\texttt{yeast\_ml8}       & 103 & 2417   & 2239   & 178  & 12.58  \\
\texttt{bankruptcy}       & 94  & 6819   & 6599   & 220  & 30.00  \\
        \bottomrule
    \end{tabular}
\end{table*}

\begin{table*}[htbp]
    \caption{Metadata for imbalanced data sets from the second experiment.}
    \label{tab:meta}
    \begin{tabular}{lrrrrrrrrrrrr}
        \toprule
Name &  $\min(\mathrm{d}_n)$ &  $\mathrm{med}(\mathrm{d}_n)$ &  $\max(\mathrm{d}_n)$ &  $\min(\mathrm{d}_p)$ &  $\mathrm{med}(\mathrm{d}_p)$ &  $\max(\mathrm{d}_p)$ &  $\min(\mathrm{d}_{np})$ &  $\mathrm{med}(\mathrm{d}_{np})$ &  $\max(\mathrm{d}_{np})$ \\
        \midrule
\texttt{arrhythmia}       & 0.79 & 2.65 & 7.10  & 0.88 & 2.43 & 3.60  & 1.02 & 2.52 & 5.70 \\
\texttt{car\_eval\_34}    & 1.41 & 2.83 & 3.46  & 1.41 & 2.45 & 3.46  & 1.41 & 2.83 & 3.46 \\
\texttt{car\_eval\_4}     & 1.41 & 2.83 & 3.46  & 1.41 & 2.45 & 3.16  & 1.41 & 2.83 & 3.46 \\
\texttt{coil\_2000}       & 0.00 & 2.06 & 4.65  & 0.00 & 2.09 & 3.89  & 0.00 & 2.12 & 4.30 \\
\texttt{isolet}           & 1.74 & 7.79 & 13.11 & 2.46 & 5.84 & 11.13 & 2.86 & 7.27 & 12.58 \\
\texttt{libras\_move}     & 0.00 & 2.88 & 5.89  & 0.71 & 2.85 & 4.69  & 0.73 & 3.11 & 5.13 \\
\texttt{oil}              & 0.13 & 1.66 & 4.75  & 0.24 & 1.55 & 2.95  & 0.17 & 1.65 & 4.37 \\
\texttt{optical\_digits}  & 0.38 & 3.13 & 4.94  & 0.49 & 2.33 & 3.99  & 0.97 & 2.93 & 4.54 \\
\texttt{ozone\_level}     & 0.33 & 2.15 & 5.35  & 0.43 & 1.38 & 3.85  & 0.38 & 1.99 & 5.35 \\
\texttt{protein\_homo}    & 0.00 & 0.78 & 4.85  & 0.00 & 1.01 & 4.11  & 0.26 & 1.08 & 5.34 \\
\texttt{satimage}         & 0.11 & 1.45 & 3.90  & 0.13 & 0.65 & 2.19  & 0.15 & 1.16 & 3.61 \\
\texttt{scene}            & 0.00 & 4.34 & 9.48  & 0.00 & 4.00 & 7.01  & 0.00 & 4.25 & 9.26 \\
\texttt{sick\_euthyroid}  & 0.00 & 2.03 & 4.95  & 0.00 & 1.44 & 3.49  & 0.02 & 1.79 & 4.93 \\
\texttt{solar\_flare\_m0} & 0.00 & 2.83 & 4.47  & 0.00 & 3.16 & 4.47  & 0.00 & 3.16 & 4.47 \\
\texttt{spectrometer}     & 0.11 & 1.20 & 6.02  & 0.14 & 3.21 & 6.66  & 0.25 & 2.17 & 7.04 \\
\texttt{thyroid\_sick}    & 0.00 & 2.45 & 5.14  & 0.02 & 1.79 & 4.09  & 0.03 & 2.10 & 4.96 \\
\texttt{us\_crime}        & 0.49 & 2.56 & 5.93  & 0.84 & 2.82 & 5.86  & 0.71 & 3.16 & 6.22 \\
\texttt{webpage}          & 1.00 & 4.69 & 12.08 & 1.00 & 4.36 & 9.17  & 0.00 & 4.58 & 11.79 \\
\texttt{yeast\_ml8}       & 0.19 & 1.84 & 2.55  & 0.98 & 1.81 & 2.44  & 0.85 & 1.83 & 2.53 \\
\texttt{bankruptcy}       & 0.07  & 1.39 & 3.61 & 0.18  & 1.42 & 3.41 & 0.18 & 1.44 & 3.65 \\
        \bottomrule
    \end{tabular}
\end{table*}

\end{document}